\documentclass[12pt]{article}

\usepackage{newtxtext,newtxmath}

\usepackage{graphicx}

\usepackage[letterpaper,margin=1in]{geometry}

\renewenvironment{abstract}
	{\quotation}
	{\endquotation}

\date{}

\makeatletter
\renewcommand{\fnum@figure}{\textbf{Figure \thefigure}}
\renewcommand{\fnum@table}{\textbf{Table \thetable}}
\makeatother

\usepackage{scicite}

\usepackage{url}

\def\scititle{
	Modern aerodynamics models do not capture important unsteady forces and delays--the failure of the quasi-steady approximation
}
\title{\bfseries \boldmath \scititle}

\author{
	Victoria~M.~Malarczyk$^{1}$,
	Marcus Hultmark$^{1\ast}$ \and
	\small$^{1}$Department of Mechanical and Aerospace Engineering, Princeton University, Princeton, NJ, 08540, USA.\and
	\small$^\ast$Corresponding author. Email: hultmark@princeton.edu
}

\begin{document} 

\maketitle


\begin{abstract} \bfseries \boldmath
Aerodynamic unsteadiness is inherent to the operation of many engineering applications, especially those that involve large-scale rotating blades, such as modern wind turbines. Over the past few decades, wind turbine rotors have grown rapidly and are now exceeding 200 meters in diameter, causing them to operate in conditions where limited empirical data are available, and where models have not been validated. Here, we use a highly pressurized wind tunnel to probe these conditions and evaluate the quasi-steady approximation, often used to simplify the modeling of the unsteady aerodynamic response of an airfoil to slow changes in inflow conditions. We find that the quasi-steady approximation is not valid for a large range of frequencies where it is normally applied. This finding suggests that the loading on wind turbine blades will be significantly underestimated when using conventional models, especially near stalling conditions, which modern wind turbines typically encounter.


\end{abstract}

\noindent

\section{Introduction}

The wind energy community has been very successful in reducing the levelized cost of wind energy over the past two decades \cite{Wiser2021, WiserBolinger2021, Bolinger2022}. There are many reasons for this trend, but an important factor is how the industry has managed to continuously increase the size of the turbines from one generation to the next \cite{Mehta2024}. This, in combination with the push for offshore wind energy, has resulted in enormous machines where all aspects of them are massive, including the tower height, blade length, and rotor diameter. The large lengthscales involved increase the individual power output of the turbine both by accessing both a larger area of the incoming wind and faster wind speeds higher up in the atmosphere. Modern wind turbine designs now use rotors on the order of several hundred meters in diameter with rated power outputs of 15 MW and higher \cite{Mehta2024}, earning them the title of the largest rotating machines ever created by mankind \cite{Veers2019}. Their massive size has brought their aerodynamics into uncharted territory and, to complicate things further, it amplifies the unsteady effects that are inherently present in wind turbine operation. Incoming winds themselves are unsteady and turbulent \cite{PorteAgel2020}, and as the wind turbine blades rotate through the wind shear of the atmospheric boundary layer, they experience unsteady inflow conditions \cite{OSTI2001}. Furthermore, any misalignment of the rotor plane with the inflow wind gives rise to effective velocity variations on the blade that affect how forces vary in time \cite{Howland2020}. The aerodynamic response of the turbine blade to these unsteady conditions have been identified as a likely cause to the rise in the fatigue and structural failure of turbine blades, disproportionately affecting large-scale blades \cite{Ashwill2009, Mishnaevsky2022}. This observation indicates a mismatch between the aerodynamic loads as they are being predicted and modeled during the design phase, and the real-world loads experienced by the turbine operating in the field. 

\begin{figure} 
	\centering
	\includegraphics[width=\textwidth]{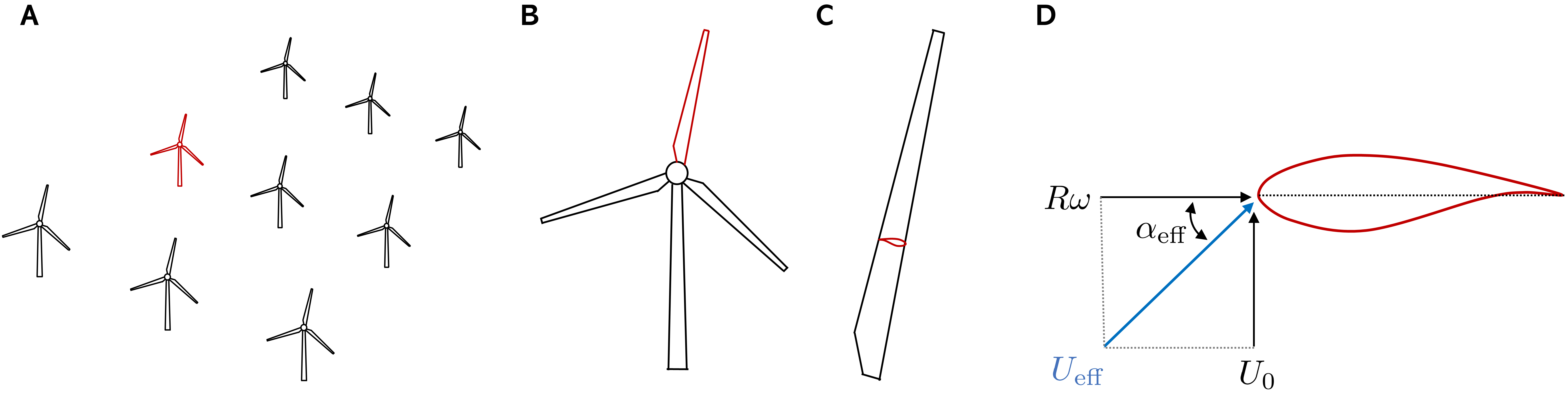}
	\caption{\textbf{Levels of abstraction associated with the scales of energy capture from the wind associated with a horizontal axis wind turbine.}
    (\textbf{A}) Wind farm. (\textbf{B}) Horizontal-axis wind turbine (HAWT). (\textbf{C}) Turbine blade. (\textbf{D}) Airfoil. The effective incoming wind speed $U_{\text{eff}}$ is shown as the vector sum of the relative velocity imparted by the blade rotation $R\omega$ ($\parallel$ to chordline) and the incoming wind speed $U_0$ ($\perp$ to chordline). The effective angle of attack $\alpha_{\text{eff}}$ is defined between $U_{\text{eff}}$ and the airfoil chordline. }
	\label{fig:abstraction}
\end{figure}

A common way to model the forces, torques, and consequent power generated by a wind turbine blade is to consider its spanwise varying cross-sectional airfoil elements as individual 2D contributors to the overall forces \cite{Hansen2015}. It has therefore become paramount for the wind turbine community to understand, and be able to accurately model, the aerodynamic response of individual airfoils to the conditions encountered by large rotors, including unsteady conditions. It is well known that unsteady effects dominate the aerodynamic response when the airfoil is operating at high angles of attack $\alpha$ (the angle between the incoming wind and the chord line of the airfoil, see Figure~\ref{fig:abstraction}D) which are near or in the separated flow region, which can lead to dynamic stall on the airfoil \cite{Shipley1994}. In order to trigger this unsteady phenomenon, the effective angle of attack of the airfoil must increase in time such that it is near or passes through the angle of attack at which the airfoil first encounters separated flow in the steady case \cite{McCroskey1981}. This angle of attack is known as the static stall angle $\alpha_{ss}$. Even if the blade has fixed pitch, Figure~\ref{fig:abstraction}D demonstrates how the change in effective angle of attack can occur from unsteadiness in the incoming windspeed $U_0$. Dynamic stall most notably involves the separation and roll-up of the boundary layer into a vortex that temporarily increases lift and delays flow separation \cite{McCroskey1981, Mulleners2012}. Given the cyclic nature of blades, as they rotate through the atmospheric wind shear, the airfoil at a given spanwise location may experience near-stalling conditions repeatedly during normal operation \cite{Shipley1995} and may trigger some or all of the stages of dynamic stall. When improperly accounted for, a large dynamic stall event is capable of setting off lasting vibrations that excite aerodynamic flutter of the blade \cite{Dunn1992, Lee1999}. In this regard, precise models of the unsteady airfoil aerodynamics are needed for a variety of operating conditions that include both fast and slow changes in the effective angle of attack of the airfoil, as they influence the aerodynamic response of the blade overall for a modern wind turbine. 


The degree of unsteadiness in the flowfield is categorized by the reduced frequency $k=\pi f c / U_0$, which is a non-dimensional parameter. Here, $f$ is the frequency of the unsteady event, $c$ is the airfoil chord length, and $U_0$ is the incoming wind speed. The reduced frequency represents the ratio of the convective time scale to the time scale of the effective angle of attack perturbation of the airfoil. Typically, for perturbations of $k > 0.05$ the flowfield response is assumed to be dynamic, such that during the progression from attached to separated flow it encounters flowfield features that cannot be captured using a quasi-steady framework \cite{Sears1941}. For $k < 0.05$, it is typically assumed that the perturbation is occurring over a long enough period such that the flowfield is able to adapt nearly instantaneously to the airfoil angle of attack as it changes in time. The flowfield in this case is often described as quasi-steady as it is assumed that the aerodynamic forces depend only on the instantaneous angle of attack of the airfoil. The $k=0.05$ cutoff value has origins in Theodorsen's foundational theory for unsteady loading in the attached flow regime, where the airfoil is pitched or heaved in low $\alpha$ conditions far from the static stall angle \cite{Theodorsen1935}. Here, it was observed that the transfer function describing phase lag and lift response, for which there is an analytical form, deviated only slightly from its steady state value for $k<0.05$ \cite{Theodorsen1935, Sears1941}. However, this value has been historically accepted as an engineering threshold with little to no physical basis, now carrying over into unsteady separated flows which include those experiencing dynamic stall, for which angles of attack are much higher by comparison and circulatory components of the lift are more critical to quantify in order to properly model the aerodynamic response. 

Many models for dynamic stall are semi-empirical in nature and rely on the quasi-steady approximation as a baseline \cite{Leishman1989, Snel1997, Oye1990, Sheng2008, Brunton2009}. Any unsteadiness that occurs on the airfoil, such as the formation of a dynamic stall vortex, is typically modeled as an additive correction to this steady baseline. For $k<0.05$, unsteady effects are assumed to be minor and, in some cases, are simply neglected altogether. Additionally, vortex formation at high $k$ has furthermore received considerably more attention for its highly identifiable signatures in measurements \cite{Dabiri2005, Mulleners2012, Eldredge2019}. By contrast, the validity of the quasi-steady approximation at low $k$ remains largely unevaluated, especially experimentally, due to the challenge associated with simultaneously matching both the flow regime of modern large-scale wind turbines and the reduced frequency where low $k$ perturbations are critical to understand and commonly encountered in daily operation. 

Besides the reduced frequency, the most important parameter governing the state of the aerodynamics of an airfoil is the chord Reynolds number, $Re_c= \frac{\rho_0 c U_0}{\mu_0}$, where $\rho_0$ is the fluid density and $\mu_0$ is the fluid viscosity. Modern offshore wind turbines operate at chord Reynolds numbers on the order of $Re_c=\mathcal{O}(10^6-10^7)$, for which there is only limited empirical data in static conditions, let alone a near complete lack of unsteady data. This is primarily due to studies of this flow regime requiring either very large wind tunnels or high freestream velocities, often on the order of 100 m/s. Large wind tunnels are costly to run, while high freestream velocities require correspondingly high pitching frequencies to match $k$, making tests infeasible in practice and limiting the parameter space that can be evaluated. The current study makes use of a pressurized wind tunnel, using the increased density of the working fluid to match the high $Re_c$ relevant to modern wind turbines, while doing so at freestream speeds less than 10 m/s. With reduced freestream speeds, lower pitching frequencies are required to achieve a given $k$, which allows for a unique investigations into unsteady phenomena in this high $Re_c$ regime, as the physical time scales of the flow vary inversely with the inflow speed. As such, compressed gas facilities are uniquely positioned to enable detailed investigations into unsteady phenomena at high $Re_c$. The focus of the present study is to use this unique testing environment to investigate the onset of unsteady behavior and evaluate the validity of the quasi-steady approximation at wind turbine relevant Reynolds numbers, over two whole decades of reduced frequencies ($0.001\leq k\leq 0.1$). 


\section{Experimental setup}

The data presented herein were collected in the High Reynolds Number Test Facility, which is a wind tunnel that can operate with air that is pressurized up to 240 bar static pressure (for more information about the facility, see Ref.~\cite{methods} and Figure~\ref{fig:setup}A). A static reference case was acquired as well as several dynamic cases, all at approximately $Re_c=6.0\times 10^6$. For the static baseline test, the wind tunnel was operating at approximately 4.5 m/s and 153 bar static gauge pressure (Table~\ref{tab:static_conditions}). For the dynamic tests, the wind tunnel was operating at approximately 6.1 m/s and 103 bar static gauge pressure (Table~\ref{tab:dynamic_conditions}). In the dynamic (unsteady) cases, the airfoil was pitched upward through a sinusoidal half-period `ramp-up' motion using a servo motor, with the mean angle $\overline\alpha$ of the profile set as the post-processed, empirically-determined static stall angle $\alpha_{ss} \approx 23^\circ$ for the tested $Re_c = 6 \times 10^6$. The mean-to-peak amplitude of the profile is $\hat\alpha = 5^\circ$, such that the aerodynamic response begins at $\alpha=18^\circ$, in an attached flow state, and finishes at $\alpha=28^\circ$, in a separated flow state, with the passage of the $\alpha_{ss}$ occurring exactly at the half-duration point of the motion profile. The motion profile was carried out as illustrated in the time series of Figure~\ref{fig:main}A.

The aerodynamic response was measured using 32 surface taps distributed about the upper and lower surfaces of a NACA0021 airfoil model. These taps were connected to corresponding differential pressure transducers situated on a printed circuit board (PCB) located inside the airfoil during test. This ensured short pressure pathways and near-instantaneous transmission of static pressure changes in the flowfield to the sensing elements on the PCB, maximizing the temporal resolution of the measurement. The instantaneous lift was computed by integrating the measured static pressure distribution about the airfoil from these taps and was nondimensionalized by the dynamic pressure of the tunnel inflow and airfoil chord length to obtain a 2D sectional lift coefficient $c_l$. Each instantaneous measurement of $c_l$ was assigned to the time stamp associated with the reading of the 32nd pressure measurement (for more information on the sensor multiplexing, see Ref.~\cite{methods}). A detailed schematic of the wind tunnel, the NACA0021 airfoil model, and the static pressure measurement system (including the PCB and surface tap locations) is shown in Figure~\ref{fig:setup}.

The NACA0021 airfoil profile used in the model is symmetric and can be considered moderately thick at 21\% chord thickness. Moderately thick airfoils are commonly used in wind turbine blade design, as they offer reliable lift curves and structural benefit, which motivated the choice of the NACA0021 for the present experiment \cite{Eppler1990, Tangler1995}. It is important to note that modern wind turbines use quite complex, proprietary airfoil geometries that vary along the span of the blade, yet the thickness of the airfoil correlates most directly to its stall and unsteady behavior. Thus, despite the detailed geometry of the airfoil not being exactly representative of what is implemented in a modern wind turbine, by choosing a more canonical airfoil profile with a relevant thickness ratio, the present study provides a more generalizable result than a specific production airfoil. 


\section{Results and Discussion}
The time histories of the lift response and the corresponding static pressure coefficients measured at each surface tap for various $k$ are shown in Figure~\ref{fig:main}B-C with the steady baseline value at each instantaneous $\alpha$ in the ramp-up motion indicated by the solid black line. The time elapsed, $t$, is reported in terms of cycle periods, $T$, where the period of a ramp-up motion, $t_r$, constitutes half a cycle $\frac{t_r}{T} = 0.5$. 
 
With the progression of the ramp-up motion, the airfoil is being subjected to a changing static pressure distribution associated with the instantaneously changing angle of attack and corresponding flowfield. For the steady baseline case (hereinafter referred to as the quasi-steady case, $k=0$), the passage of $\alpha_{ss}$ indicates the condition at which the adverse pressure gradient on the upper surface of the airfoil is severe enough such that the boundary layer has insufficient momentum to follow the airfoil curvature and separates from the surface. This separated flow state, often referred to as the stalled flow state, is associated with the abrupt drop in lift observed for the quasi-steady case in Figure~\ref{fig:main}B. 

\begin{figure} 
	\centering
	\includegraphics[width=0.6\textwidth]{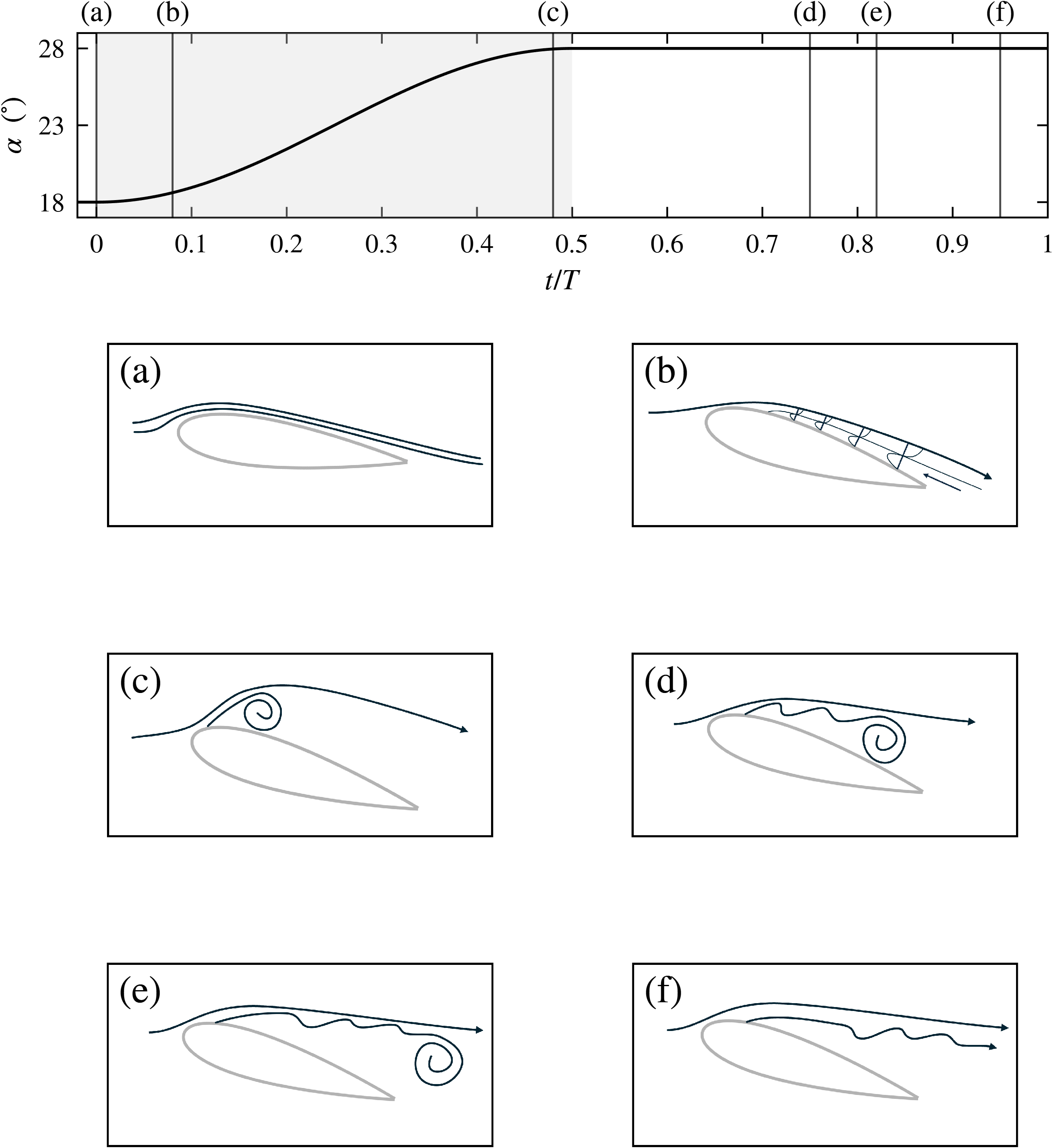}
	\caption{\textbf{Key stages of dynamic stall development.}
   Illustrations of the flowfield for (\textbf{A}) Attached flow. (\textbf{B}) Flow reversal. (\textbf{C}) Vortex formation.
   (\textbf{D}) Vortex convection. (\textbf{E}) Vortex shed. (\textbf{F}) Separated flow. These illustrations are considered representative only for a highly unsteady flow, such as $k = 0.1$. The time history of the angle of attack variation for the ramp-up motion is shown on top. Time stamps on this plot are labelled to correspond to the illustrations of \textbf{A}-\textbf{F} to indicate where each stage may be encountered during the motion. Actual timestamps for a given motion as they relate to dynamic stall development are highly $k$-dependent. }
	\label{fig:dynstall}
\end{figure}

One of the strongest indicators of unsteadiness for a given $k$ is the low-pressure signature associated with the dynamic stall vortex. Figure~\ref{fig:dynstall} illustrates the six stages associated with the dynamic stall \cite{McCroskey1981, Mulleners2012}. As the ramp-up motion progresses, the separated shear layer forming from the movement of the boundary layer separation point toward the leading edge wraps into a vortex through the growth of a shear instability as the flow near the airfoil surface is reversed. As the vortex takes shape, its associated suction signature increases, observed in the tap static pressure coefficient reading shown in Figure~\ref{fig:main}C. The maximum suction value, across all taps, is achieved near the leading edge (typically tap 2) with elevated suction observed in the chordwise extent of the airfoil stretching to the mid chord (taps 1-10). The chordwise extent of this suction signature has been interpreted as the chordwise extent of the vortex \cite{Kiefer2022}. At a later time, after the peak suction is achieved, the elevated suction signature is observed to move downstream passing over taps that are located closer to the trailing edge (taps 11-16). This behavior is likely associated with the convection of the vortex downstream. Interestingly, it is only for the the higher values of $k$ that the downstream convection of the elevated suction signature is clearly observed. Specifically, in the vicinity of taps 11-15, an elevated suction bump downstream of the peak suction at the leading edge can be seen at $t/T \approx 0.8$ for $k=0.1$, at $t/T \approx 0.55$ for $k=0.05$, and at $t/T \approx 0.42$ for $k=0.02$, and $t/T \approx 0.36$ for $k=0.01$. For $k \leq 0.002$, there does not appear to be an observable elevated suction signature downstream of the suction peak, indicating that for these $k$ values dynamic stall likely had not completely entered the vortex formation stage, or at least is very weak. For these low $k$ values, any unsteadiness observed in the static pressure readings may be related to boundary layer behaviors arising during the flow reversal stage.

Comparing the dynamic data ($k\neq0$) to the quasi-steady data ($k=0$), the aerodynamic response looks remarkably different for all tested $k$. The lowest reduced frequency tested, $k=0.001$, represents a 50-fold decrease below the $k$ that is commonly assumed to be slow enough to warrant a quasi-steady treatment ($k_{\text{cutoff}}=0.05$). It is clear that the findings herein do not support such an approximation, with an elevated lift peak and a significant stall delay present after passage of the static stall angle for all test cases. This finding discloses that the aerodynamic response possesses a high degree of sensitivity to $k$. Furthermore, the stall mechanism for the lowest reduced frequency tested, $k=0.001$, appears to evolve in a remarkably different way compared to the steady baseline, with elevated loading spread out over a longer period of time (cf. Figure~\ref{fig:main}B).

\section{Conclusion}

\begin{figure} 
	\centering
	\includegraphics[width=0.75\textwidth]{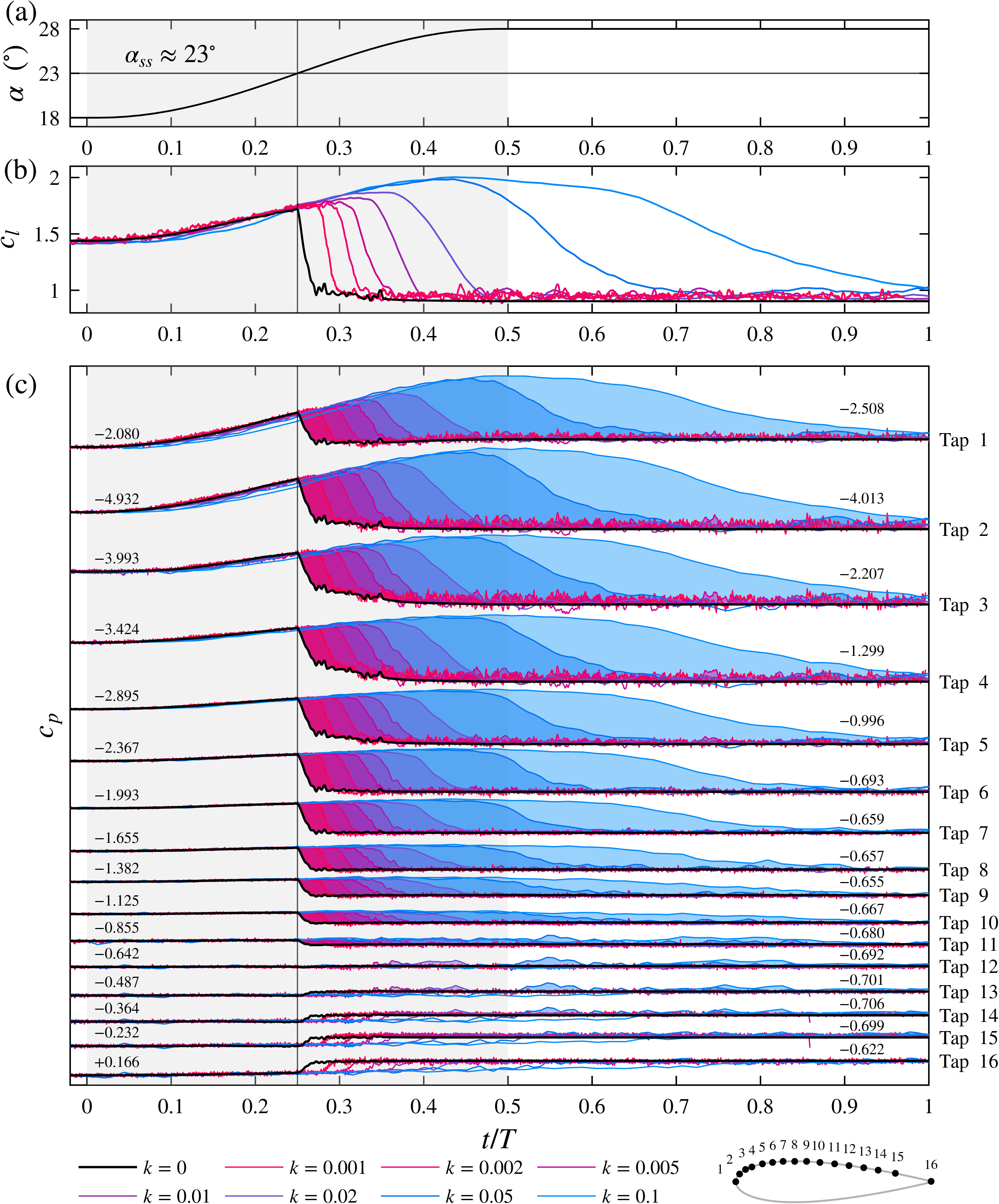}
	\caption{\textbf{Unsteady aerodynamic response to airfoil ramp-up motions of various $k$.} Time histories of the (\textbf{A}) Angle of attack variation performed by the airfoil during the ramp-up motion, (\textbf{B}) Sectional lift coefficient, (\textbf{C}) Static pressure coefficients as measured from individual surface taps on the upper (suction) surface at $Re_c = 6 \times 10^6$. Tap number on the right corresponds to the chordwise location indicated on the airfoil inset below. The curves are shifted vertically for clarity, with the labelled values near $t/T \approx 0$ and $t/T \approx 0.9$ indicating the true $c_p$ magnitude experienced at each time stamp respectively. Legend colors apply to both \textbf{B} and \textbf{C}, where the steady baseline is reported as the $k = 0$ line. The corresponding-color shading under each curve in \textbf{C} is used to emphasize the elevated suction associated with unsteadiness in the aerodynamic response.}
	\label{fig:main}
\end{figure}

This study shows that there is significant unsteadiness present in the aerodynamic response of an airfoil that is being pitched past its static stall angle, even at reduced frequencies much slower than the accepted $k_{\text{cutoff}}=0.05$ threshold that is commonly used a cutoff value for implementation of the quasi-steady approximation. For the approximation to be valid at a given $k$, the curves shown in Figure~\ref{fig:main}B-C should all collapse atop the $k=0$ case. It is evident that it is not the case for any of the $k$ of the tested motions, and the quasi-steady approximation fails for all.

Measurements of the static pressure distribution show definitive low-pressure signatures of both dynamic stall vortex formation and downstream convection in response to ramp-up motions down to $k=0.005$, which is ten times slower than $k_{\text{cutoff}}=0.05$. The typical unsteady aerodynamic modeling approach assumes that any unsteadiness in the response can reasonably be neglected if the dynamic stall vortex is not fully formed. For the lowest $k$ motions studied herein, it is inconclusive to determine whether the suction at the leading edge may be attributed to the complete formation of a dynamic stall vortex due to its weaker peak suction signature and lack of a corresponding downstream convection signature. Nonetheless, this finding still represents a non-negligible unsteadiness in the form of both a stall delay and an elevated leading edge suction signature that must be accounted for in order to accurately evaluate the unsteady loading on airfoil sections for a modeled wind turbine blade at very low $k$ maneuvers. This suggests that for wind turbines of growing size, even slow, low frequency motions can produce elevated lift forces that complicate modeling and design efforts. 

A wind turbine blade effectively never operates in perfectly steady wind conditions. Our finding presents a conundrum for any approach that attempts to simplify the modeling of unsteady load development about an airfoil for slow motions, as this common approximation appears to overlook significant physics that can cause forces, torques and power to go unmodeled. This will evidently contribute to unanticipated failures and higher maintenance costs for wind turbine blades specifically as they are being designed for virtually non-existent, simplified hypothetical operating conditions. However, it should also be pointed out that this finding presents an opportunity as, after all, increased forces and power output from a system that benefits from just that can be highly beneficial, but only if it can be predicted and accounted for in its design. 

\newpage

\begin{figure} 
	\centering
	\includegraphics[width=1\textwidth]{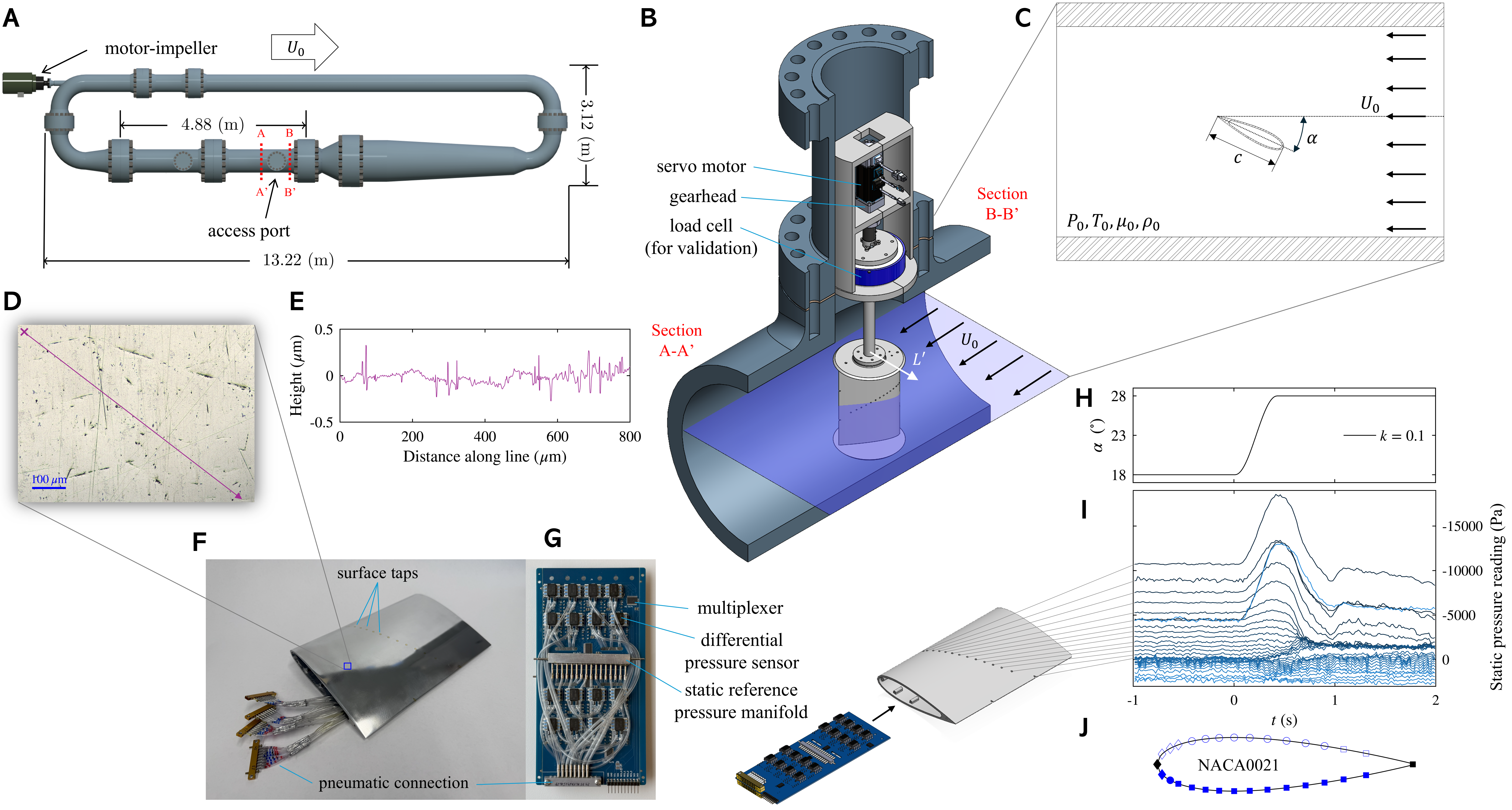}
	\caption{\textbf{Schematic of the experimental setup and wind tunnel facility.}
        (\textbf{A}) Schematic of the High Reynolds number Test Facility (HRTF) at Princeton University. Flow direction of freestream is emphasized by the arrow labelled $U_0$. (\textbf{B}) Cutaway of the HRTF access port with the airfoil setup inserted. (\textbf{C}) Schematic of the test environment with the freestream velocity $U_0$, chord length $c$, and angle of attack $\alpha$ labelled. The tunnel working fluid properties are indicated with the `0' subscript. (\textbf{D}) Confocal microscope image of the airfoil model surface with scale bar. (\textbf{E}) Characterization of the roughness height along the magenta line in \textbf{D} in the direction indicated by the arrow starting at `X.' (\textbf{F}) Photo of the hand-polished NACA0021 airfoil model with pneumatic connectors to the PCB exposed. (\textbf{G}) Image showing the printed circuit board (PCB) with 32 differential pressure sensors and a reference static pressure manifold. (\textbf{H}) Ramp-up profile in terms of angle of attack $\alpha$ as performed by the servo motor for $k=0.1$. (\textbf{I}) Representative static pressure measurements obtained from the surface taps on the airfoil model. (\textbf{J}) Chordwise tap locations on the NACA0021 airfoil model. Open symbols indicate the upper (suction) surface, and filled symbols indicate lower (pressure) surface. The symbols correspond to specific sensor ratings in order to maximize measurement resolution: $\lozenge \pm 400$ mbar, $\bigcirc \pm 100$ mbar, $\square \pm 40$ mbar. }
	\label{fig:setup}
\end{figure}

\clearpage 

%
\bibliography{bibliography} 
\bibliographystyle{sciencemag}

%
%
%
%
%
%


\section*{Acknowledgments}
The authors would like to thank Dan Hoffman for help with the airfoil setup and operation of the pressurized wind tunnel facility.
 
\paragraph*{Funding:}
The authors gratefully acknowledge funding from the Army Research Office through Grant W911NF-22-1-0187 and the New Jersey Wind Institute Fellowship Program.

\paragraph*{Author contributions:}
Conceptualization, M.H. and V.M.; methodology, V.M.; software, V.M.; experiments, V.M.; data analysis, V.M.; writing, V.M.; and funding acquisition, M.H. and V.M.

\paragraph*{Competing interests:}
There are no competing interests to declare.

\paragraph*{Data and materials availability:} Both the dynamic (unsteady) and static baseline data are publicly available at https://doi.org/10.34770/h9sj-fk79 .

\subsection*{Supplementary materials}
Materials and Methods\\
Figures S1 to S4\\
Tables S1 to S4\\
Caption for Data S1


\newpage


\renewcommand{\thefigure}{S\arabic{figure}}
\renewcommand{\thetable}{S\arabic{table}}
\renewcommand{\theequation}{S\arabic{equation}}
\renewcommand{\thepage}{S\arabic{page}}
\setcounter{figure}{0}
\setcounter{table}{0}
\setcounter{equation}{0}
\setcounter{page}{1} 


\begin{center}
\section*{Supplementary Materials for\\ \scititle}

Victoria~M.~Malarczyk,
Marcus~Hultmark$^\ast$\\ 
\small$^\ast$Corresponding author. Email: hultmark@princeton.edu\\
    
\end{center}

\subsubsection*{This PDF file includes:}
Materials and Methods\\
Figures S1 to S4\\
Tables S1 to S4\\
Caption for Data S1

\newpage


\subsection*{Materials and Methods}

The unique test conditions presented in this study ($Re_c = 6 \times 10^6$ and $0.001 \leq k \leq 0.1$) were made possible through use of the High Reynolds number Test Facility (HRTF) located at the Gas Dynamics Laboratory at Princeton University. A schematic of the HRTF is shown in Figure~\ref{fig:setup}A. The HRTF is a closed-loop, recirculating, high pressure wind tunnel that uses dried compressed air as its working fluid and supports static pressures of up to 240 bar. The facility supports freestream velocities of up to 8.5 m/s resulting in $Ma$ $\leq$ 0.03, well below the incompressible limit. The pressurization of the working fluid allows for near-uniform mean freestream velocity across the tunnel cross section, and turbulence levels of less than 1$\%$ for all $Re_c$ studied. The test section of the tunnel is circular in cross section, measuring 0.49 m in diameter and 4.88 m in length. The NACA0021 airfoil model is inserted into the test section using an access port that is located 1.1 m downstream of the test section entrance. Once the model is inserted, the tunnel is sealed and pressurized, and all signals fed to and read from the model must be transmitted through a feedthrough cable. The facility is described in further detail in Ref. \cite{jimenez2010intermediate}. 

\subsubsection*{Tunnel diagnostic capability}
The tunnel inflow conditions were measured from a host of sensors located inside the HRTF during operation. These were processed by two National Instruments data acquisition cards read in by a PC. The freestream velocity was measured using a Pitot-static tube (United Sensor model USNH-A-38) that was fixed upstream of the test section. The readings from this sensor were read through a differential pressure transducer (Validyne DP-15) with a range of $\pm$2 psi. The tunnel static temperature was measured using a resistance temperature detector (RTD, Omega Technologies Corporation) located upstream of the tunnel’s flow-conditioning section through a plug. The tunnel static pressure was measured using a transducer (Omega model PX419) connected to a static pressure port also located through a plug upstream of the tunnel’s flow-conditioning section. 

For the range of pressures that the working fluid (air) is compressed to, the compressibility factor, $Z$, only changes by 10$\%$ at its most extreme, indicating that the working fluid behaves as an ideal gas over most of the states that it is being tested at in the HRTF \cite{Zagarola1996}. As such, the working fluid density could be computed from the static tunnel pressure and temperature measurements using the ideal gas law. The procedure for determining dynamic viscosity at pressure in the HRTF is detailed in Ref. \cite{Zagarola1996} for which there is a necessary correction.

\subsubsection*{NACA0021 airfoil model}

A NACA0021 airfoil model was custom-built to evaluate high $Re_c$ aerodynamics. The chord length of the model is $c$ = 0.17 m with aspect ratio AR = 1.5. The model is equipped with circular end plates that are 0.190 m in diameter. Both the airfoil model and the endplates were manufactured from a single block of Aluminum 6061 using electrical discharge machining. The model surface polished to a mirror finish by hand with total roughness between $0.24 \times 10^{-6} \leq \kappa / c \leq 0.94 \times 10^{-6}$. 


A high-torque servo motor (Applied Motion model J0400-351-4-000) executed the angle of attack changes for the experiment and was attached to the airfoil setup on one end. The motor has a maximum continuous torque rating of 1.27 N-m with a maximum rated speed of 3000 rpm. This continuous torque was enhanced using a 25:1 reduction gearhead to provide over 30 N-m of torque. For the static tests, the angle of attack was changed using a stream command function in the proprietary servo drive software (SVX Servo Suite) which was able to match the desired angle of attack position within 0.0014$^\circ$. For the dynamic tests, the motor was hardware-controlled using an analog input provided by a Keysight 33500B Waveform function generator. The motor was able to consistently follow any dynamic profile that it was sent within 10 encoder counts or 0.014$^\circ$. 

A support structure was created to withstand both radial and thrust loads to sufficiently transmit the rotational motion from the servo motor to the airfoil about a singular central axis. This structure primarily holds the servo motor which is connected to a force-torque sensor (JR3) via rigid shaft coupler. A small hole in the support structure allows for a shaft that is connected to the opposite side of the force-torque sensor to pass through and rotate about its symmetric axis without interference, to which the airfoil with endplates is attached on its end. The force-torque sensor supports ranges $\pm$200 N and $\pm$25 N-m and was used to verify the integrated force and moment measurements from the static pressure measurement system. It is important to emphasize that all integrated aerodynamic force and moment measurements reported herein originate from the static pressure measurement system detailed in the following section. The complete airfoil assembly weighs 11.4 kg (25.1 lbs) without instrumentation.

\subsubsection*{Digital static pressure measurement system}

\begin{figure} 
	\centering
	\includegraphics[width=0.65\textwidth]{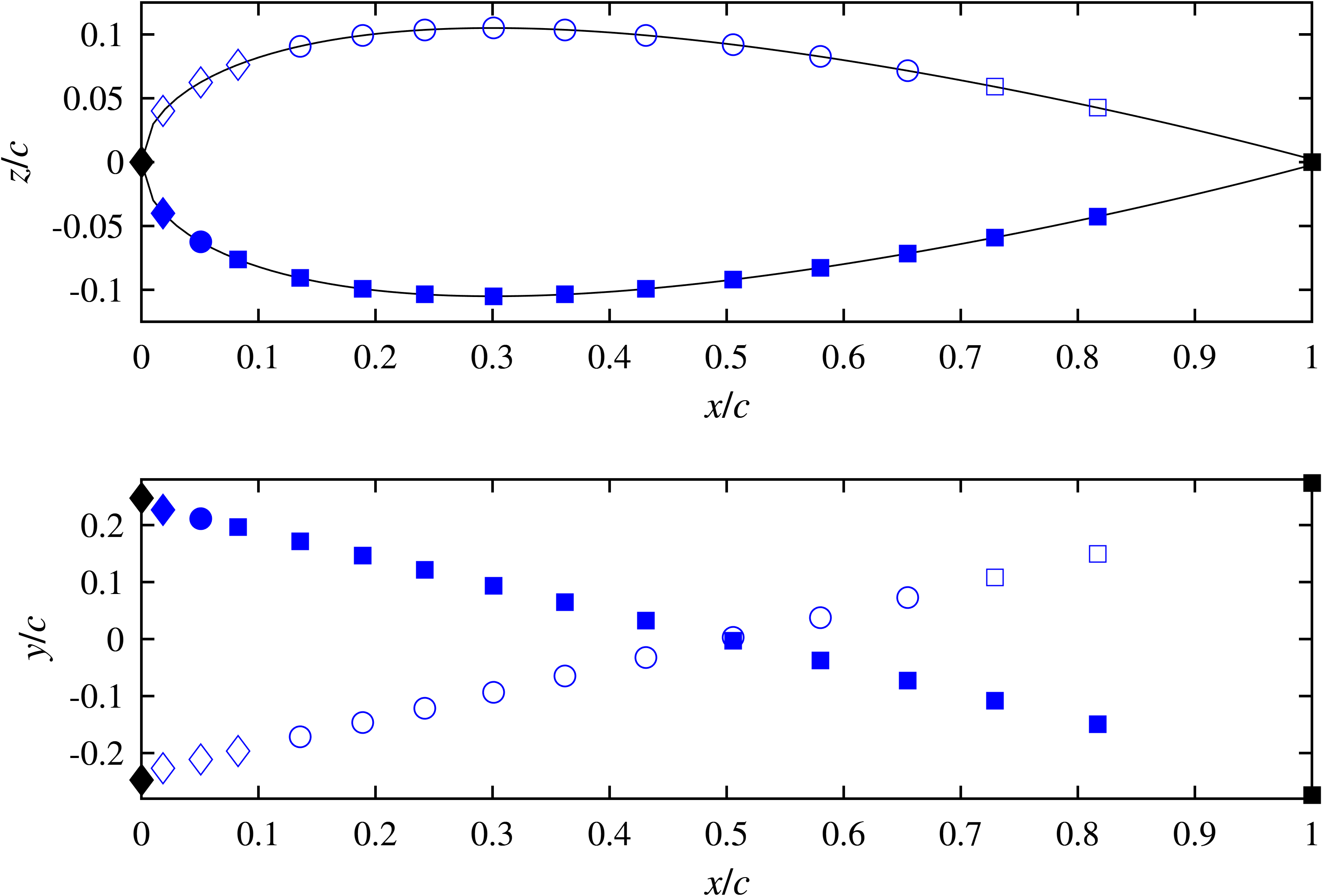}
	\caption{\textbf{Locations of the surface taps on the NACA0021 airfoil model. }
	The distribution of the tap locations in the x-coordinate (chordwise), y-coordinate (spanwise), and z-coordinate (thicknesswise). All dimensions are normalized by chord length, 0.170 m. The marker shapes at each location correspond to a different pressure sensor resolution: $\lozenge \pm 400$ mbar, $\bigcirc \pm 100$ mbar, $\square \pm 40$ mbar. The black markers correspond to the leading edge and trailing edge taps.}
	\label{fig:supp_locations} 
\end{figure}

The presented results hinge on accurate measurement of the static pressure distribution about the airfoil. The static pressure was measured using 32 sensing holes (`surface taps') distributed along the upper and lower surfaces of the airfoil. Each sensing hole is 0.5 mm in diameter, and bored into a 3 mm diameter brass insert. The brass inserts were press-fit into 32 corresponding 3 mm diameter holes on the airfoil model and sanded down to restore the airfoil curvature. Each brass insert is connected to a urethane tube that routes the static pressure experienced at the airfoil surface directly to one port of a dual barbed port Honeywell TruStability HSC Series differential pressure sensor. To minimize pressure lag, the differential pressure sensors are mounted on a printed circuit board (PCB) that is located inside the airfoil during test. The other barbed port of the differential pressure sensor is therefore connected to a manifold that is open to the static tunnel pressure, which equilibrates throughout the volume of fluid inside of the airfoil model by means of several openings located along the central axis of the setup. Each sensor corresponds to a particular sensing range (either 400 mbar, 100 mbar, or 40 mbar) to maximize the resolution of the local surface pressure measurement. The sensing range also corresponds to the chordwise tap location on the airfoil surface for the model, with higher ranges located toward the leading edge (see Figure~\ref{fig:supp_locations}).

The differential pressure sensors were configured for digital I$^2$C communication protocol for use with a Raspberry Pi. Additionally, the readings were multiplexed because it was not possible to achieve simultaneous readings for all 32 sensors as all were manufactured with the same default sensor address. Instead, the sensor address was `reassigned' through use of four 8-channel multiplexers, where each of the 32 sensors were attached to their own channel on a multiplexer whose address could be user-defined. The multiplexers switch between the 32 sensors and transmit the reading from one sensor at a time to then be recorded by the Raspberry Pi. The measurements were taken sequentially from tap 1 to 32 which repeated until the sampling window was terminated.  For the purpose of computing the lift coefficient in this study, the time stamp of the 32nd sensor for each repetition of the multiplexing sequence was assigned to the integrated lift value to give an approximate instantaneous time stamp for the loading as measured by the 32-sensor sweep. The multiplexer operated at 2 kHz resulting in individual pressure sensors being sampled at 2 kHz/32 = 62.5 Hz.

\subsubsection*{Experimental procedure}

For the static test, used to determine the static baseline value, the airfoil model was held fixed at a given angle of attack $\alpha$, starting from $\alpha = -5 ^\circ$ and increasing to $\alpha = 40 ^\circ$. The angles of attack near the critical stalling angle were sampled at 0.1$^\circ$ increments in order to properly capture the stalling process. For each $\alpha$, a few seconds of pause was held before static pressure measurements were taken to ensure that any unsteadiness from changing angles had convected from the airfoil. Table~\ref{tab:static_conditions} indicates the number of convective time scales over which the measurement was taken. The captured static pressure distribution was then time-averaged to give a representative result for a given $\alpha$. This pressure distribution could then be nondimensionalized using the measured dynamic pressure from the tunnel diagnostics. The tunnel conditions were captured at the start and end of the angle of attack sweep using an automated LabVIEW script with an acquisition frequency of 5 kHz and the averaged value between the two captures was used for the nondimensionalization. The difference between the starting and ending tunnel conditions was less than 5$\%$. These averaged tunnel conditions for the static airfoil data are reported in Supplement T\ref{tab:static_conditions}. For the $-5^\circ \leq \alpha \leq 40^\circ$ sweep, the blockage for the model in the tunnel is between $6.7\% \leq BR(\alpha) \leq 17.1\%$, as demonstrated in Figure~\ref{fig:setup}C.

For the dynamic tests, the airfoil model was pitched upward (in a `ramp-up' motion) using a servo motor such that its angle of attack $\alpha$ varied in a sinusoidal motion. The ramp-up motion was performed as,

\begin{equation}
\label{eq:rampup}
\alpha(t) =
\begin{cases}
\bar{\alpha} - \hat{\alpha}, & t \leq 0, \\[6pt]
\bar\alpha + \hat\alpha \text{sin}(\omega t - \frac{\pi}{2}), &  0 \leq t \leq t_r = \frac{1}{2f_s}\\[6pt]
\bar{\alpha} + \hat{\alpha}, & t \geq t_r, \\[6pt]
\end{cases}
\end{equation}

The mean angle of attack $\bar \alpha$ for the profile was taken to be approximately the static stall angle which was experimentally found to be $\alpha_{ss} \approx 23^\circ$ for the given airfoil profile, surface roughness, inflow turbulence, and $Re_c$. 

The unsteadiness in the flow response to the ramp-up motion is typically characterized using the reduced frequency $k$ typically defined as, 
\begin{equation}
\label{eq:reduced_frequency}
k = \frac{\omega b}{U_0} = \frac{\pi f_sc}{U_0}
\end{equation}
where $U_0$ is the free stream velocity, $c$ is the chord length, $b = \frac{c}{2}$ is the length in semichords, and $\omega = 2\pi f_s$ is the pitching frequency in rad/second while $f_s$ is the pitching frequency expressed in terms of cycles/second or Hz. The ramp-up motion in this experiment represents a half-sinusoidal cycle, such that the motion occurs over a period $t_r$ that is half the period of the continuous sinusoidal oscillation, or $t_r = \frac{1}{2 f_s} = \frac{T}{2}$. The reduced frequency $k$ in terms of ramp-up motion parameters can be expressed as,
\begin{equation}
k = \frac{ \pi c}{2t_rU_0} 
\end{equation}
This is consistent with the periods of the ramp-up motions expressed in Table~\ref{tab:dynamic_conditions}.

\subsubsection*{Data processing}
\begin{figure} 
	\centering
	\includegraphics[width=0.75\textwidth]{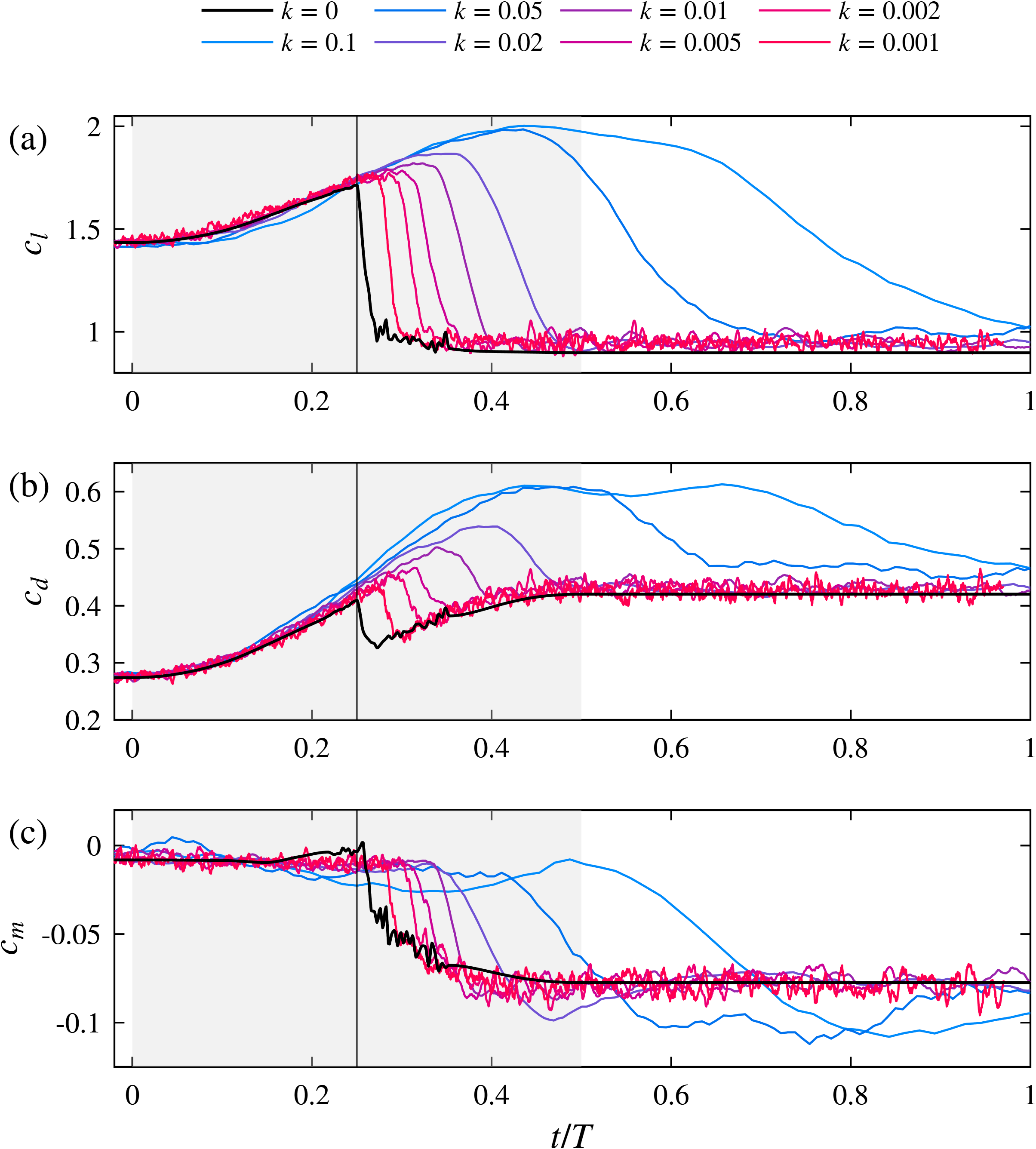}
	\caption{\textbf{Evolution of unsteady aerodynamic response to various $k$. }
		 Time histories for the (\textbf{A}) Lift coefficient. (\textbf{B}) Pressure drag coefficient. (\textbf{C}) Quarter-chord moment coefficient. Legend applies to all plotted lines.}
	\label{fig:supp_forces} 
\end{figure}

The forces and quarter-chord moment (see Figure~\ref{fig:supp_forces}) experienced by the airfoil model are integrated measurements of its static pressure distribution. The lift is the component of the force perpendicular to the relative inflow velocity, while the drag component lies parallel to the relative inflow velocity. Given the two-dimensional nature of the airfoil profile, the resultant lift $L'$, form drag $D'$, and quarter-chord moment $M_{c/4}'$ are reported in terms of unit span, indicated by the apostrophe. Nondimensionalized by the dynamic pressure $q_0$ and chord length $c$, the coefficients will be expressed as,
\begin{equation}
    c_l = \frac{L'}{q_0 c} \qquad
    c_d = \frac{D'}{q_0 c} \qquad 
    c_{m,c/4} = \frac{M_{c/4}'}{q_0 c^2} \qquad
    \label{eq:s4}
\end{equation}

Given the finite number of tap locations, a discrete midpoint integration scheme with the panel method was used. As an intermediate step, the two-dimensional normal force coefficient $c_n$ (perpendicular to chord length), axial force coefficient $c_a$ (parallel to chord length), and leading edge moment coefficient $c_{m, \text{LE}}$ (computed about the leading edge instead of the quarter-chord point) were computed as follows,
\begin{equation}
c_n =
\frac{1}{c}
\sum_{i=1}^{15}
\left(
c_{p,l}\!\left(x_{i+\frac{1}{2}}\right)
-
c_{p,u}\!\left(x_{i+\frac{1}{2}}\right)
\right)
\Delta x_i
\end{equation}
\begin{equation}
c_a =
\frac{1}{c}
\sum_{i=1}^{15}
\left(
c_{p,u}\!\left(x_{i+\frac{1}{2}}\right)\frac{dy_{u}\left(x_{i+\frac{1}{2}}\right)}{dx_{i+\frac{1}{2}}}
-
c_{p,l}\!\left(x_{i+\frac{1}{2}}\right)\frac{dy_{l}\left(x_{i+\frac{1}{2}}\right)}{dx_{i+\frac{1}{2}}}
\right)
\Delta x_i
\end{equation}
\begin{equation}
c_{m, \text{LE}} =
\frac{1}{c^2}
\sum_{i=1}^{15}
\left(
\left(c_{p,u} - c_{p,l}\right)x_{i+\frac{1}{2}}
+ 
c_{p,u}\frac{dy_{u}\left(x_{i+\frac{1}{2}}\right)}{dx_{i+\frac{1}{2}}}y_{u, i+\frac{1}{2}} 
-
c_{p,l}\frac{dy_{l}\left(x_{i+\frac{1}{2}}\right)}{dx_{i+\frac{1}{2}}}y_{l, i+\frac{1}{2}}
\right) 
\Delta x_i
\end{equation}

\noindent where $x_{i+\frac{1}{2}}$ denotes the center of panel $i$ and $\Delta x_i = x_{i+1} - x_i$ indicates the length of the panel over the dimensional chordwise coordinate. The subscript `u' refers to the upper, suction surface of the airfoil and the subscript `l' refers to the lower, pressure surface of the airfoil. The $y_u$ coordinate is directed positive above the $x$ axis and $y_l$ coordinate is directed negative below the $x$ axis. 

The lift coefficient $c_l$ and form drag coefficient $c_d$ can therefore be found through rotating the normal and axial forces through the angle of attack $\alpha$, 
\begin{equation}
c_l = c_n \text{cos}(\alpha) - c_a \text{sin}(\alpha)
\end{equation}
\begin{equation}
c_d = c_n \text{sin}(\alpha) + c_a \text{cos}(\alpha)
\end{equation}
The quarter-chord moment coefficient $c_{m, c/4}$ was found from shifting the point about which the moment is calculated from the leading edge to the quarter chord point,
\begin{equation}
c_{m, c/4} = c_{m, \text{LE}} + \frac{1}{4}c_l
\end{equation}

To present a self-contained study, $c_d$ Figure~\ref{fig:supp_forces}B reports only the pressure (form) drag component as can be calculated from the static pressure distribution. All experimental results and their discussion presented herein is based on entirely uncorrected data. Only the aforementioned processing was applied, with no additional considerations of blockage or three-dimensional effects.

\begin{figure} 
	\centering
	\includegraphics[width=0.75\textwidth]{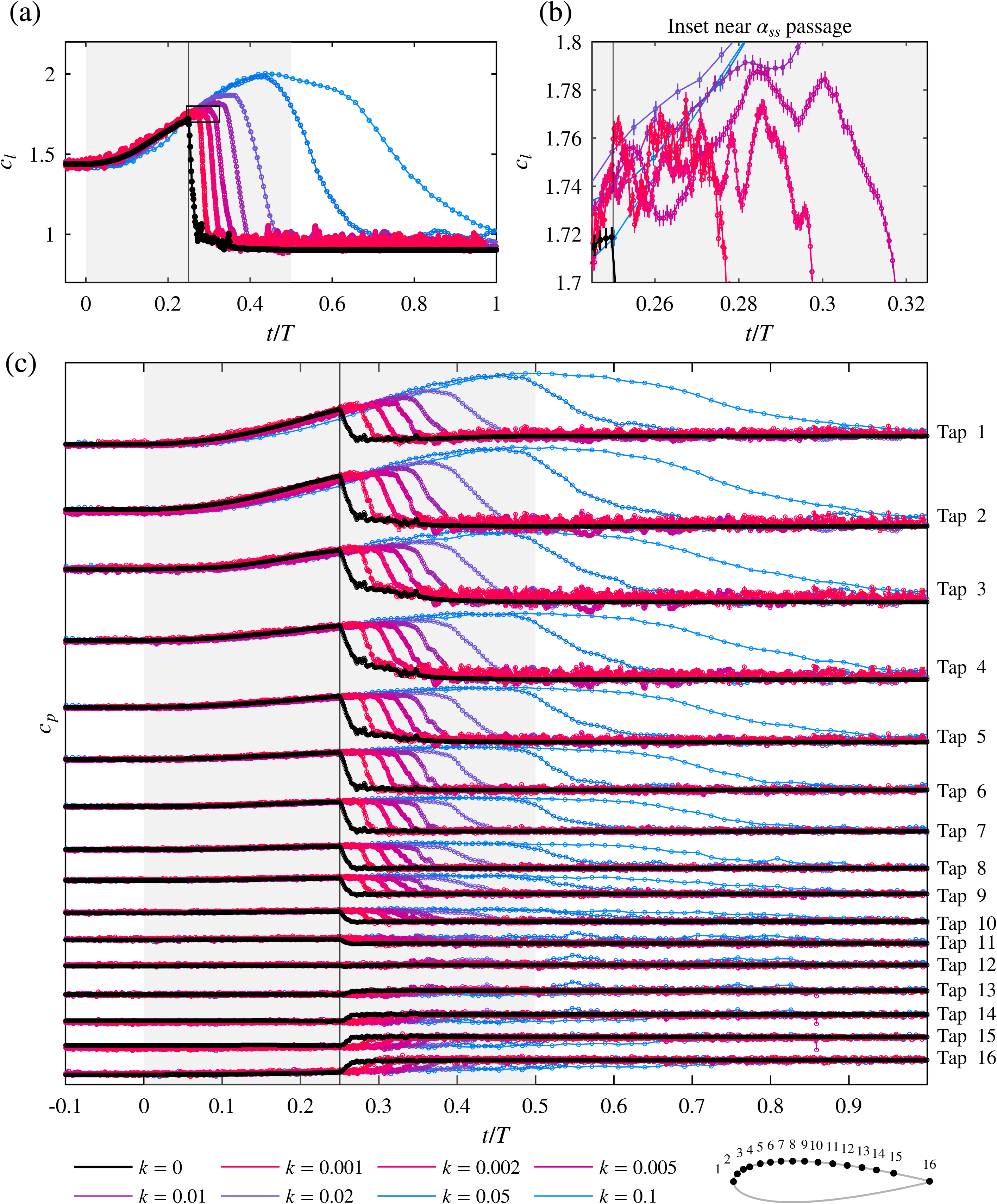}
	\caption{\textbf{Propagated error for data. }
		Error bars indicate the propagated sensor error for (\textbf{A}) lift coefficient, (\textbf{B}) inset of lift coefficient near passage of $\alpha_{ss}$, and (\textbf{C}) static pressure coefficients. If the error bar is not shown it is understood that the marker is larger than the error associated with that data point. The associated sensor uncertainties are summarized in Table~\ref{tab:supp_uncertainties}. }
	\label{fig:supp_errorbars} 
\end{figure}

\begin{figure} 
	\centering
	\includegraphics[width=0.75\textwidth]{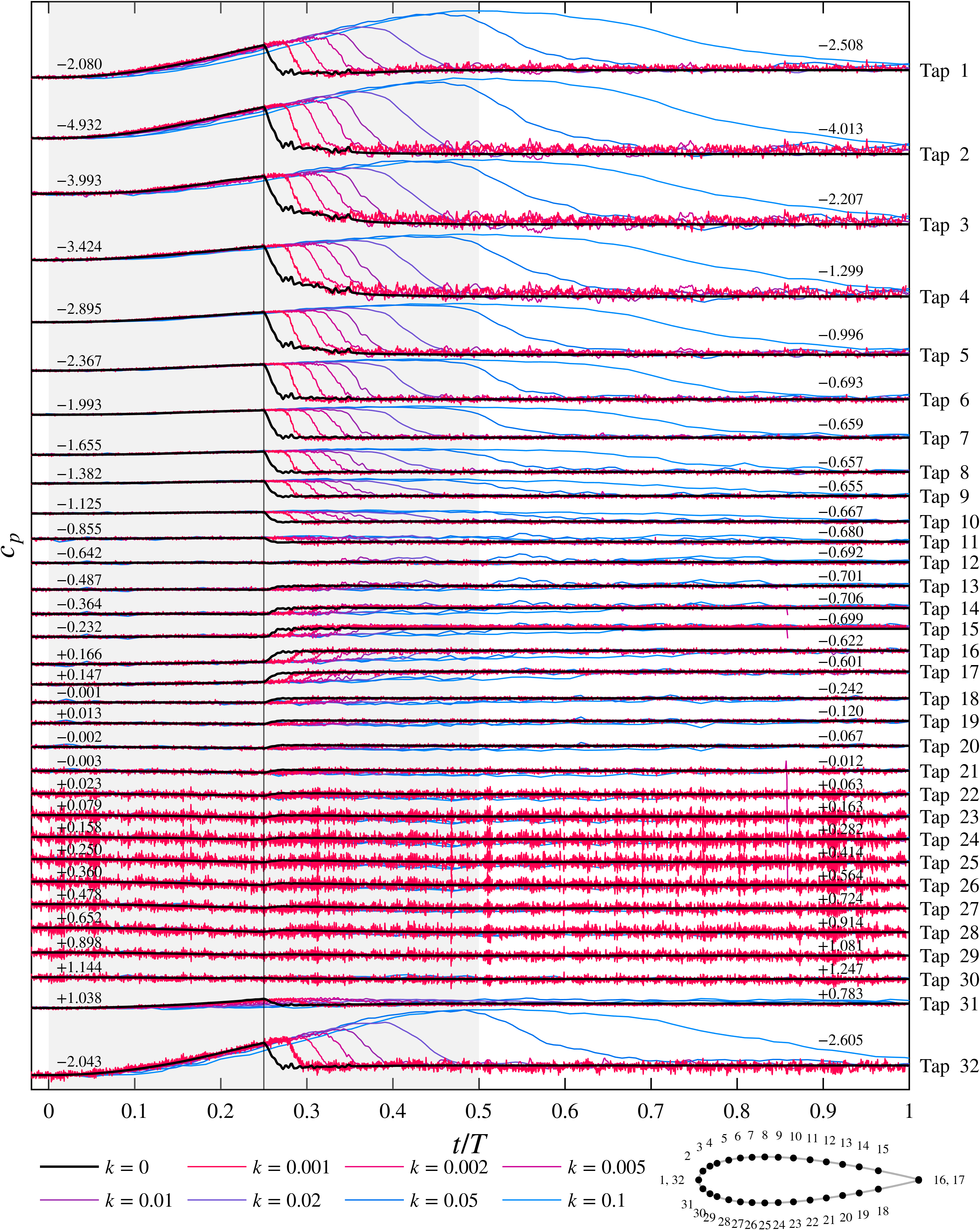}
	\caption{\textbf{Time histories of static pressure measurements at all 32 chordwise surface tap locations. }
		 Static pressure measurements are reported in their nondimensionalized form as coefficients. Legend applies to all plotted lines. The NACA0021 airfoil inset in the lower right indicates the chordwise locations of the 32 surface taps where measurements were taken. Taps 1 $\&$ 32 are both located at the leading edge, and taps 16 $\&$ 17 are both located at the trailing edge.}
	\label{fig:supp_32sensors} 
\end{figure}

\newpage

\begin{table}[t]
\centering
\caption{Summary of sensor uncertainties. The sensor uncertainties related to static differential pressure refer to the differential pressure sensors on the PCB inside the airfoil model.}
\label{tab:supp_uncertainties}

\begin{tabular}{ccc}
\hline
Parameter & Uncertainty Symbol & Value \\
\hline
Working fluid temperature & $\Delta T_0$ & $\pm\text{ }0.15\text{ }K$ \\
Working fluid static pressure & $\Delta P_0 $& $\pm\text{ }1.0\text{ }\%$ \\
Working fluid density & $\Delta \rho_0$ & $\pm\text{ }0.036\text{ }\%$ \\
Working fluid dynamic viscosity  & $\Delta \mu_0$ &  $\pm\text{ }0.8\text{ }\%$ \\
Dynamic pressure & $\Delta q_0$ &  $\pm\text{ }0.184\text{ }\%$ \\
Reynolds number & $\Delta Re_c$ &  $\pm\text{ }0.805\text{ } \%$ \\
Angle of attack & $\Delta \alpha$ &  $\pm\text{ }0.0014 ^\circ$ \\
Static differential pressure ($\pm$ 40 mbar) & $\Delta P_s$ & $\pm\text{ }0.1 \text{ mbar } (10 \text{ Pa}) $ \\
Static differential pressure ($\pm$ 100 mbar) & $\Delta P_s$ & $\pm\text{ } 0.25\text{ mbar } (25\text{ Pa})$  \\
Static differential pressure ($\pm$ 400 mbar) & $\Delta P_s$ & $\pm\text{ } 1\text{ mbar } (100 \text{ Pa}) $  \\
\hline
\end{tabular}
\end{table}

\begin{table}[htbp]
\centering
\caption{Definition of variables for tunnel conditions listed in Tables~\ref{tab:static_conditions} and~\ref{tab:dynamic_conditions} .
    }
\label{tab:tunnel_var}

\begin{tabular}{ll}
\\
\hline
Symbol & Description \\
\hline
$Re_c$ & Chord Reynolds number\\
$U_0$  & Freestream velocity of the wind tunnel inflow \\
$\rho_0$ & Air density of the working fluid \\
$P_0$ & Static pressure of the working fluid \\
$T_0$ &  Temperature of the working fluid \\
$\mu_0$ & Dynamic viscosity of the working fluid \\
$k$ &  Reduced frequency of the ramp-up motion \\
$f_s$ &  Dimensional frequency of the sinusoidal input, such that $T = \frac{1}{f_s}$ \\
$t_r$ &  Period of a single ramp motion, such that $T = 2t_r$ \\
\hline
\end{tabular}
\end{table}

\newpage 

\begin{table} 
	\centering
	\caption{\textbf{Tunnel conditions for static baseline $k=0$.} The time over which the static result was averaged for each $\alpha$ is given in the column indicated by $\Delta t^*$, reported in terms of convective time scales, $t^* = tU_0/c$.}
	\label{tab:static_conditions} 

\begin{tabular}{cccccccc}
\hline
$Re_c$ & $U_0$ (m/s) & $\rho_0$ (kg/m$^3$) & $P_0$ (bar [psi]) & $T_0$ ($^\circ$C) & $\mu_0$ (Pa$\cdot$s) & $\Delta t^*$ \\
\hline
6.02$ \times 10^6$ & 4.51 & 174.6 & 153.3 [2224] & 30.6 & 2.224$\times 10^{-5}$ & 797 \\
\hline
\end{tabular}
\end{table}

\begin{table} 
	\centering
	\caption{\textbf{Tunnel conditions for dynamic (unsteady) tests at various $k \neq 0$. } The number of cycles over which the dynamic result was phase-averaged is given by the column `No. tests.'
		}
	\label{tab:dynamic_conditions} 

\small
\begin{tabular}{cccccccccc}
\hline
$k$ & $Re_c$ & $U_0$ (m/s) & $\rho_0$ (kg/m$^3$) & $P_0$ (bar [psi]) & $T_0$ ($^\circ$C) & $\mu_0$ (Pa$\cdot$s) & $f_s$ (Hz) & $t_r$ (s) & No. Tests \\
\hline
0.001 & 6.07$\times 10^6$ & 6.18 & 118.9 & 101.4 [1471] & 28.3 & 2.058$\times 10^{-5}$ & 0.012 & 42.7 & 22 \\
0.002 & 6.08$\times 10^6$ & 6.07 & 121.4 & 103.3 [1498] & 27.6 & 2.061$\times 10^{-5}$ & 0.023 & 22.0 & 12 \\
0.005 & 6.04$\times 10^6$ & 6.08 & 121.0 & 104.0 [1508] & 29.9 & 2.071$\times 10^{-5}$ & 0.057 & 8.7 & 20 \\
0.01 & 6.07$\times 10^6$ & 6.09 & 121.4 & 104.1 [1510] & 29.5 & 2.070$\times 10^{-5}$ & 0.114 & 4.4 & 20 \\
0.02 & 6.09$\times 10^6$ & 6.09 & 121.4 & 103.6 [1502] & 28.1 & 2.064$\times 10^{-5}$ & 0.228 & 2.2 & 20 \\
0.05 & 6.19$\times 10^6$ & 6.06 & 122.3 & 101.5 [1472] & 21.5 & 2.035$\times 10^{-5}$ & 0.567 & 0.9 & 8 \\
0.1 & 6.11$\times 10^6$ & 6.06 & 121.7 & 102.3 [1484] & 24.8 & 2.049$\times 10^{-5}$ & 1.137 & 0.4 & 20 \\
\hline
\end{tabular}

\end{table}


\clearpage 

\paragraph{Caption for Data S1.}
\textbf{Data for the static and dynamic (unsteady) airfoil tests.}
This data set contains the static pressure coefficient measurements from the 32 surface taps and the corresponding integrated lift, drag, and moment coefficient values experimentally obtained by the NACA0021 airfoil model under both static and dynamic (unsteady) conditions. These experiments were performed at high Reynolds number and low Mach number. This data was acquired in the High Reynolds number Test Facility at Princeton University, which is a closed-loop wind tunnel that can be pressurized up to 3500 psi. 



\end{document}